\documentclass[12pt]{iopart}%

\begin{document}
\title[The Cosmological Constant]{Extracting the Cosmological Constant from the Wheeler DeWitt Equation in a
Modified Gravity Theory}
\author{Remo Garattini}

\begin{abstract}
We discuss how to extract information about the cosmological constant from the
Wheeler-DeWitt equation, considered as an eigenvalue of a Sturm-Liouville
problem. A generalization to a $f\left(  R\right)  $ theory is taken under
examination. The equation is approximated to one loop with the help of a
variational approach with Gaussian trial wave functionals. We use a zeta
function regularization to handle with divergences. A renormalization
procedure is introduced to remove the infinities together with a
renormalization group equation.

\end{abstract}

\address{Universit\`{a} degli Studi di Bergamo, Facolt\`{a} di Ingegneria,\\ Viale
Marconi 5, 24044 Dalmine (Bergamo) ITALY.\\INFN - sezione di Milano, Via Celoria 16, Milan, Italy}
\ead{remo.garattini@unibg.it}

\section{Introduction}

The Einstein's field equations represent a fundamental set of information
regarding the laws of space-time. They are represented by%
\begin{equation}
R_{\mu\nu}-\frac{1}{2}g_{\mu\nu}R+\Lambda_{c}g_{\mu\nu}=\kappa T_{\mu\nu},
\label{ein}%
\end{equation}
where $T_{\mu\nu}$ is the energy-momentum tensor of some matter fields,
$\kappa=8\pi G$ with $G$ the Newton's constant and $\Lambda_{c}$ is the
cosmological constant . The sourceless version of Eqs.$\left(  \ref{ein}%
\right)  $ is simply%
\begin{equation}
R_{\mu\nu}-\frac{1}{2}g_{\mu\nu}R+\Lambda_{c}g_{\mu\nu}=0. \label{eins}%
\end{equation}
It is well known that there exists a huge discrepancy between the
observed\cite{Lambda} and the computed value of the cosmological constant. It
amounts approximately to a factor of 120 orders of magnitude: this is the
\textit{cosmological constant problem}. One possible approach to such a
problem is given by the Wheeler-DeWitt equation (WDW)\cite{De Witt}. The WDW
equation can be extracted from the Einstein's field equations with and without
matter fields in a very simple way. If we introduce a time-like unit vector
$u^{\mu}$ such that $u\cdot u=-1$, then after a little rearrangement, we
get\footnote{See Ref.\cite{Remo} for more details.}%
\begin{equation}
\mathcal{H=}\left(  2\kappa\right)  G_{ijkl}\pi^{ij}\pi^{kl}-\frac{\sqrt{g}%
}{2\kappa}\!{}\!\left(  \,^{3}R-2\Lambda_{c}\right)  =0.
\end{equation}
$^{3}R$ is the scalar curvature in three dimensions. This is the time-time
component of Eqs.$\left(  \ref{eins}\right)  $. It represents a constraint at
the classical level and the invariance under \textit{time} reparametrization.
Its quantum counterpart%
\begin{equation}
\mathcal{H}\Psi\mathcal{=}0 \label{WDW}%
\end{equation}
is the WDW equation. If we integrate over the hypersurface $\Sigma$ and we
define%
\begin{equation}
\hat{\Lambda}_{\Sigma}=\left(  2\kappa\right)  G_{ijkl}\pi^{ij}\pi^{kl}%
-\frac{\sqrt{g}}{2\kappa}\!{}\,^{3}R,
\end{equation}
Eq.$\left(  \ref{WDW}\right)  $ can be cast into the following form%
\begin{equation}
\int\mathcal{D}\left[  g_{ij}\right]  \Psi^{\ast}\left[  g_{ij}\right]
\left[  \int_{\Sigma}d^{3}x\hat{\Lambda}_{\Sigma}\right]  \Psi\left[
g_{ij}\right]  =-\frac{\Lambda_{c}}{\kappa}V\int\mathcal{D}\left[
g_{ij}\right]  \Psi^{\ast}\left[  g_{ij}\right]  \Psi\left[  g_{ij}\right]  ,
\end{equation}
where we have multiplied Eq.$\left(  \ref{WDW}\right)  $ by $\Psi^{\ast
}\left[  g_{ij}\right]  $, we have functionally integrated over the three
spatial metric $g_{ij}$ and where we have defined the volume of the
hypersurface $\Sigma$ as $V=\int_{\Sigma}d^{3}x\sqrt{g}$. Thus one can
formally re-write the WDW equation as%
\begin{equation}
\frac{1}{V}\frac{\int\mathcal{D}\left[  g_{ij}\right]  \Psi^{\ast}\left[
g_{ij}\right]  \int_{\Sigma}d^{3}x\hat{\Lambda}_{\Sigma}\Psi\left[
g_{ij}\right]  }{\int\mathcal{D}\left[  g_{ij}\right]  \Psi^{\ast}\left[
g_{ij}\right]  \Psi\left[  g_{ij}\right]  }=\frac{1}{V}\frac{\left\langle
\Psi\left\vert \int_{\Sigma}d^{3}x\hat{\Lambda}_{\Sigma}\right\vert
\Psi\right\rangle }{\left\langle \Psi|\Psi\right\rangle }=-\frac{\Lambda_{c}%
}{\kappa}, \label{WDW2}%
\end{equation}
We can gain more information considering a separation of the spatial part of
the metric into a background term, $\bar{g}_{ij}$, and a quantum fluctuation,
$h_{ij}$,%
\begin{equation}
g_{ij}=\bar{g}_{ij}+h_{ij}.
\end{equation}
Thus Eq.$\left(  \ref{WDW2}\right)  $ becomes%
\begin{equation}
\frac{1}{V}\frac{\left\langle \Psi\left\vert \int_{\Sigma}d^{3}x\left[
\hat{\Lambda}_{\Sigma}^{\left(  0\right)  }+\hat{\Lambda}_{\Sigma}^{\left(
1\right)  }+\hat{\Lambda}_{\Sigma}^{\left(  2\right)  }+\ldots\right]
\right\vert \Psi\right\rangle }{\left\langle \Psi|\Psi\right\rangle }%
=-\frac{\Lambda_{c}}{\kappa}\Psi\left[  g_{ij}\right]  , \label{WDW3}%
\end{equation}
where $\hat{\Lambda}_{\Sigma}^{\left(  i\right)  }$ represents the $i^{th}$
order of perturbation in $h_{ij}$. By observing that the kinetic part of
$\hat{\Lambda}_{\Sigma}$ is quadratic in the momenta, we only need to expand
the three-scalar curvature $\int d^{3}x\sqrt{g}{}\ ^{3}R$ up to the quadratic
order and we get%
\[
\int_{\Sigma}d^{3}x\sqrt{\bar{g}}\left[  -\frac{1}{4}h\triangle h+\frac{1}%
{4}h^{li}\triangle h_{li}-\frac{1}{2}h^{ij}\nabla_{l}\nabla_{i}h_{j}%
^{l}+\right.
\]%
\begin{equation}
\left.  +\frac{1}{2}h\nabla_{l}\nabla_{i}h^{li}-\frac{1}{2}h^{ij}R_{ia}%
h_{j}^{a}+\frac{1}{2}hR_{ij}h^{ij}+\frac{1}{4}h\left(  R^{\left(  0\right)
}\right)  h\right]  \label{rexp}%
\end{equation}
where $h$ is the trace of $h_{ij}$ and $R^{\left(  0\right)  }$ is the three
dimensional scalar curvature on-shell. Eq.$\left(  \ref{WDW2}\right)  $
represents the Sturm-Liouville problem associated with the cosmological
constant. The related boundary conditions are dictated by the choice of the
trial wavefunctionals which, in our case are of the Gaussian type. Different
types of wavefunctionals correspond to different boundary conditions.
Extracting the TT tensor contribution from Eq.$\left(  \ref{WDW2}\right)  $
approximated to second order in perturbation of the spatial part of the metric
into a background term, $\bar{g}_{ij}$, and a perturbation, $h_{ij}$, we get%
\begin{equation}
\hat{\Lambda}_{\Sigma}^{\bot}=\frac{1}{4V}\int_{\Sigma}d^{3}x\sqrt{\bar{g}%
}G^{ijkl}\left[  \left(  2\kappa\right)  K^{-1\bot}\left(  x,x\right)
_{ijkl}+\frac{1}{\left(  2\kappa\right)  }\left(  \triangle_{2}\right)
_{j}^{a}K^{\bot}\left(  x,x\right)  _{iakl}\right]  . \label{p22}%
\end{equation}
Here $G^{ijkl}$ represents the inverse DeWitt metric and all indices run from
one to three. The propagator $K^{\bot}\left(  x,x\right)  _{iakl}$ can be
represented as
\begin{equation}
K^{\bot}\left(  \overrightarrow{x},\overrightarrow{y}\right)  _{iakl}:=%
{\displaystyle\sum_{\tau}}
\frac{h_{ia}^{\left(  \tau\right)  \bot}\left(  \overrightarrow{x}\right)
h_{kl}^{\left(  \tau\right)  \bot}\left(  \overrightarrow{y}\right)
}{2\lambda\left(  \tau\right)  }, \label{proptt}%
\end{equation}
where $h_{ia}^{\left(  \tau\right)  \bot}\left(  \overrightarrow{x}\right)  $
are the eigenfunctions of $\triangle_{2}$, whose explicit expression for the
massive case will be shown in the next section. $\tau$ denotes a complete set
of indices and $\lambda\left(  \tau\right)  $ are a set of variational
parameters to be determined by the minimization of Eq.$\left(  \ref{p22}%
\right)  $. The expectation value of $\hat{\Lambda}_{\Sigma}^{\bot}$ is easily
obtained by inserting the form of the propagator into Eq.$\left(
\ref{p22}\right)  $ and minimizing with respect to the variational function
$\lambda\left(  \tau\right)  $. Thus the total one loop energy density for TT
tensors becomes%
\begin{equation}
\frac{\Lambda}{8\pi G}=-\frac{1}{4}%
{\displaystyle\sum_{\tau}}
\left[  \sqrt{\omega_{1}^{2}\left(  \tau\right)  }+\sqrt{\omega_{2}^{2}\left(
\tau\right)  }\right]  . \label{1loop}%
\end{equation}
The above expression makes sense only for $\omega_{i}^{2}\left(  \tau\right)
>0$, where $\omega_{i}$ are the eigenvalues of $\triangle_{2}$.

\section{One loop energy Regularization and Renormalization for a $f\left(
R\right)  =R$ theory}

The Spin-two operator for the Schwarzschild metric in the Regge and Wheeler
representation\cite{Regge Wheeler}, leads to the following system of equations
$\left(  r\equiv r\left(  x\right)  \right)  $%
\begin{equation}
\left\{
\begin{array}
[c]{c}%
\left[  -\frac{d^{2}}{dx^{2}}+\frac{l\left(  l+1\right)  }{r^{2}}+m_{1}%
^{2}\left(  r\right)  \right]  f_{1}\left(  x\right)  =\omega_{1,l}^{2}%
f_{1}\left(  x\right) \\
\\
\left[  -\frac{d^{2}}{dx^{2}}+\frac{l\left(  l+1\right)  }{r^{2}}+m_{2}%
^{2}\left(  r\right)  \right]  f_{2}\left(  x\right)  =\omega_{2,l}^{2}%
f_{2}\left(  x\right)
\end{array}
\right.  , \label{p34}%
\end{equation}
where reduced fields have been used and the proper geodesic distance from the
\textit{throat} of the bridge has been considered. Close to the throat, the
effective masses are%
\begin{equation}
\left\{
\begin{array}
[c]{c}%
m_{1}^{2}\left(  r\right)  \simeq-m_{0}^{2}\left(  M\right) \\
\\
m_{2}^{2}\left(  r\right)  \simeq m_{0}^{2}\left(  M\right)
\end{array}
\right.  ,
\end{equation}
where we have defined a parameter $r_{0}>2MG$ and $m_{0}^{2}\left(  M\right)
=3MG/r_{0}^{3}$. The main reason for introducing a new parameter resides in
the fluctuation of the horizon that forbids any kind of approach. It is now
possible to explicitly evaluate Eq.$\left(  \ref{1loop}\right)  $ in terms of
the effective mass. To further proceed we use the W.K.B. method used by `t
Hooft in the brick wall problem\cite{tHooft} and we count the number of modes
with frequency less than $\omega_{i}$, $i=1,2$. Thus the one loop total energy
for TT tensors becomes%
\begin{equation}
\frac{\Lambda}{8\pi G}=\rho_{1}+\rho_{2}=-\frac{1}{16\pi^{2}}\sum_{i=1}%
^{2}\int_{\sqrt{m_{i}^{2}\left(  r\right)  }}^{+\infty}\omega_{i}^{2}%
\sqrt{\omega_{i}^{2}-m_{i}^{2}\left(  r\right)  }d\omega_{i},
\label{tote1loop}%
\end{equation}
where we have included an additional $4\pi$ coming from the angular
integration. Here, we use the zeta function regularization method to compute
the energy densities $\rho_{1}$ and $\rho_{2}$. Note that this procedure is
completely equivalent to the subtraction procedure of the Casimir energy
computation where the zero point energy (ZPE) in different backgrounds with
the same asymptotic properties is involved. To this purpose, we introduce the
additional mass parameter $\mu$ in order to restore the correct dimension for
the regularized quantities. Such an arbitrary mass scale emerges unavoidably
in any regularization scheme. One gets%
\begin{equation}
\rho_{i}\left(  \varepsilon\right)  =\frac{m_{i}^{4}\left(  r\right)  }%
{256\pi^{2}}\left[  \frac{1}{\varepsilon}+\ln\left(  \frac{\mu^{2}}{m_{i}%
^{2}\left(  r\right)  }\right)  +2\ln2-\frac{1}{2}\right]  , \label{zeta1}%
\end{equation}
$i=1,2$. The renormalization is performed via the absorption of the divergent
part into the re-definition of the bare classical constant $\Lambda$
\begin{equation}
\Lambda\rightarrow\Lambda_{0}+\Lambda^{div}.
\end{equation}

The remaining finite value for the cosmological constant reads%
\[
\frac{\Lambda_{0}}{8\pi G}=\frac{1}{256\pi^{2}}\left\{  m_{1}^{4}\left(
r\right)  \left[  \ln\left(  \frac{\mu^{2}}{\left\vert m_{1}^{2}\left(
r\right)  \right\vert }\right)  +2\ln2-\frac{1}{2}\right]  \right.
\]

\begin{equation}
\left.  +m_{2}^{4}\left(  r\right)  \left[  \ln\left(  \frac{\mu^{2}}%
{m_{2}^{2}\left(  r\right)  }\right)  +2\ln2-\frac{1}{2}\right]  \right\}
=\left(  \rho_{1}\left(  \mu\right)  +\rho_{2}\left(  \mu\right)  \right)
=\rho_{eff}^{TT}\left(  \mu,r\right)  . \label{lambda0}%
\end{equation}

The quantity in Eq.$\left(  \ref{lambda0}\right)  $ depends on the arbitrary
mass scale $\mu.$ It is appropriate to use the renormalization group equation
to eliminate such a dependence. To this aim, we impose that\cite{RGeq}%

\begin{equation}
\frac{1}{8\pi G}\mu\frac{\partial\Lambda_{0}^{TT}\left(  \mu\right)
}{\partial\mu}=\mu\frac{d}{d\mu}\rho_{eff}^{TT}\left(  \mu,r\right)  .
\label{rg}%
\end{equation}
Solving it we find that the renormalized constant $\Lambda_{0}$ should be
treated as a running one in the sense that it varies provided that the scale
$\mu$ is changing%

\begin{equation}
\Lambda_{0}\left(  \mu,r\right)  =\Lambda_{0}\left(  \mu_{0},r\right)
+\frac{G}{16\pi}\left(  m_{1}^{4}\left(  r\right)  +m_{2}^{4}\left(  r\right)
\right)  \ln\frac{\mu}{\mu_{0}}. \label{lambdamu}%
\end{equation}
Substituting Eq.$\left(  \ref{lambdamu}\right)  $ into Eq.$\left(
\ref{lambda0}\right)  $ we find%
\begin{equation}
\frac{\Lambda_{0}\left(  \mu_{0},M\right)  }{8\pi G}=-\frac{1}{128\pi^{2}%
}\left\{  m_{0}^{4}\left(  M\right)  \left[  \ln\left(  \frac{m_{0}^{2}\left(
M\right)  }{4\mu_{0}^{2}}\right)  +\frac{1}{2}\right]  \right\}  .
\label{lambdamu0}%
\end{equation}
Eq.$\left(  \ref{lambdamu0}\right)  $ has a maximum
when\textbf{\footnote{\textbf{Remark }Note that in any case, the maximum of
$\Lambda$ corresponds to the minimum of the energy density.}}%
\begin{equation}
\frac{1}{e}=\frac{m_{0}^{2}\left(  M\right)  }{4\mu_{0}^{2}}\qquad
\Longrightarrow\qquad\Lambda_{0}\left(  \mu_{0}\right)  =\frac{Gm_{0}%
^{4}\left(  M\right)  }{32\pi}=\frac{G\mu_{0}^{4}}{2\pi e^{2}}.
\label{LambdansM}%
\end{equation}
The computed cosmological constant appears to depend on the Schwarzschild
radius. This dependence simply reflects the fact that the chosen background
introduce one physical scale: the Schwarzschild radius. Nothing prevent us to
consider a more general situation where the scalar curvature $R$ is replaced
by a generic function of $R$. Therefore, we will consider the Sturm-Liouville
problem of Eq.$\left(  \ref{WDW2}\right)  $ in the context of a $f\left(
R\right)  $ theory\footnote{A recent review on the problem of $f\left(
R\right)  $ theories can be found in Ref.\cite{CF}. A more general discussion
on modified gravities of the type $f\left(  R\right)  ,f\left(  G\right)  $
and $f\left(  R,G\right)  $ where $G$ is the Gauss-Bonnet invariant, can be
found in Ref.\cite{NO}.}.

\section{One loop energy Regularization and Renormalization for a generic
$f\left(  R\right)  $ theory in a Hamiltonian formulation}

In this section, we report the main steps discussed in Ref.\cite{CG} for a
$f\left(  R\right)  $ theory in connection with the Sturm-Liouville problem of
Eq.$\left(  \ref{WDW2}\right)  $. Although a $f\left(  R\right)  $ theory does
not need a cosmological constant, rather it should explain it, we shall
consider the following Lagrangian density describing a generic $f(R)$ theory
of gravity
\begin{equation}
\mathcal{L}=\sqrt{-g}\left(  f\left(  R\right)  -2\Lambda\right)
,\qquad{with}\;f^{\prime\prime}\neq0,\label{lag}%
\end{equation}
where $f\left(  R\right)  $ is an arbitrary smooth function of the scalar
curvature and primes denote differentiation with respect to the scalar
curvature. A cosmological term is added also in this case for the sake of
generality, because in any case, Eq.$\left(  \ref{lag}\right)  $ represents
the most general lagrangian to examine. Obviously $f^{\prime\prime}=0$
corresponds to GR. The generalized Hamiltonian density for the $f\left(
R\right)  $ theory assumes the form\footnote{See Ref.\cite{Querella} for
technical details.}%
\[
\mathcal{H}=\frac{1}{2\kappa}\left[  -\sqrt{g}f^{\prime}\left(  R\right)
\left(  {}^{\left(  3\right)  }R-2\Lambda_{c}-3K_{ij}K^{ij}+K^{2}\right)
\right.
\]%
\begin{equation}
\left.  +V(\mathcal{P})+2g^{ij}\left(  \sqrt{g}f^{\prime}\left(  R\right)
\right)  _{\mid ij}-2p^{ij}K_{ij}\right]  ,\label{Hamf(R)}%
\end{equation}
where%
\begin{equation}
\mathcal{P=}-6\sqrt{g}f^{\prime}\left(  R\right)
\end{equation}
and%
\begin{equation}
V(\mathcal{P})=\sqrt{g}\left[  Rf^{\prime}\left(  R\right)  -f\left(
R\right)  \right]  .\label{V(P)}%
\end{equation}
Henceforth, the superscript $3$ indicating the spatial part of the metric will
be omitted on the metric itself. When $f\left(  R\right)  =R$, $V(\mathcal{P}%
)=0$ as it should be. Eq.$\left(  \ref{Hamf(R)}\right)  $ becomes%
\[
\mathcal{H}=f^{\prime}\left(  R\right)  \left[  \left(  2\kappa\right)
G_{ijkl}\pi^{ij}\pi^{kl}{}-\frac{\sqrt{g}}{2\kappa}{}\left(  ^{\left(
3\right)  }R-2\Lambda_{c}\right)  \right]
\]%
\begin{equation}
+\frac{1}{2\kappa}\left[  \sqrt{g}f^{\prime}\left(  R\right)  \left(
2K_{ij}K^{ij}\right)  +V(\mathcal{P})+2g^{ij}\left(  \sqrt{g}f^{\prime}\left(
R\right)  \right)  _{\mid ij}-2p^{ij}K_{ij}\right]  .
\end{equation}
Since%
\begin{equation}
p^{ij}=\sqrt{g}K^{ij},
\end{equation}
then we obtain%
\[
\mathcal{H}=f^{\prime}\left(  R\right)  \left[  \left(  2\kappa\right)
G_{ijkl}\pi^{ij}\pi^{kl}{}-\frac{\sqrt{g}}{2\kappa}{}\left(  ^{\left(
3\right)  }R-2\Lambda_{c}\right)  \right]
\]%
\begin{equation}
+\frac{1}{2\kappa}\left[  2\sqrt{g}K_{ij}K^{ij}\left(  f^{\prime}\left(
R\right)  -1\right)  +V(\mathcal{P})+2g^{ij}\left(  \sqrt{g}f^{\prime}\left(
R\right)  \right)  _{\mid ij}\right]
\end{equation}
and transforming into canonical momenta, one gets%
\[
\mathcal{H}=f^{\prime}\left(  R\right)  \left[  \left(  2\kappa\right)
G_{ijkl}\pi^{ij}\pi^{kl}{}-\frac{\sqrt{g}}{2\kappa}{}\left(  ^{\left(
3\right)  }R-2\Lambda_{c}\right)  \right]
\]%
\begin{equation}
+2\left(  2\kappa\right)  \left[  G_{ijkl}\pi^{ij}\pi^{kl}+\frac{\pi^{2}}%
{4}\right]  \left(  f^{\prime}\left(  R\right)  -1\right)  +\frac{1}{2\kappa
}\left[  V(\mathcal{P})+2g^{ij}\left(  \sqrt{g}f^{\prime}\left(  R\right)
\right)  _{\mid ij}\right]  .\label{Hamf(R)_1}%
\end{equation}
By imposing the Hamiltonian constraint, we obtain
\[
f^{\prime}\left(  R\right)  \left[  \left(  2\kappa\right)  G_{ijkl}\pi
^{ij}\pi^{kl}{}-\frac{\sqrt{g}}{2\kappa}{}^{\left(  3\right)  }R\right]
{}+2\left(  2\kappa\right)  \left[  G_{ijkl}\pi^{ij}\pi^{kl}+\frac{\pi^{2}}%
{4}\right]  \left(  f^{\prime}\left(  R\right)  -1\right)
\]%
\begin{equation}
+\frac{1}{2\kappa}\left[  V(\mathcal{P})+2g^{ij}\left(  \sqrt{g}f^{\prime
}\left(  R\right)  \right)  _{\mid ij}\right]  =-f^{\prime}\left(  R\right)
\sqrt{g}\frac{\Lambda_{c}}{\kappa}%
\end{equation}
If we assume that $f^{\prime}\left(  R\right)  \neq0$ the previous expression
becomes%
\[
\left[  \left(  2\kappa\right)  G_{ijkl}\pi^{ij}\pi^{kl}{}-\frac{\sqrt{g}%
}{2\kappa}{}^{\left(  3\right)  }R\right]  +\left(  2\kappa\right)  \left[
G_{ijkl}\pi^{ij}\pi^{kl}+\frac{\pi^{2}}{4}\right]  \frac{2\left(  f^{\prime
}\left(  R\right)  -1\right)  }{f^{\prime}\left(  R\right)  }%
\]%
\begin{equation}
+\frac{1}{2\kappa f^{\prime}\left(  R\right)  }\left[  V(\mathcal{P}%
)+2g^{ij}\left(  \sqrt{g}f^{\prime}\left(  R\right)  \right)  _{\mid
ij}\right]  =-\sqrt{g}\frac{\Lambda_{c}}{\kappa}{}.
\end{equation}
Now, we integrate over the hypersurface $\Sigma$ to obtain%
\begin{equation}
\int_{\Sigma}d^{3}x\left\{  \left[  \left(  2\kappa\right)  G_{ijkl}\pi
^{ij}\pi^{kl}{}-\frac{\sqrt{g}}{2\kappa}{}^{\left(  3\right)  }R\right]
+\left(  2\kappa\right)  \left[  G_{ijkl}\pi^{ij}\pi^{kl}+\frac{\pi^{2}}%
{4}\right]  \frac{2\left(  f^{\prime}\left(  R\right)  -1\right)  }{f^{\prime
}\left(  R\right)  }\right\}
\end{equation}%
\begin{equation}
+\int_{\Sigma}d^{3}x\frac{1}{2\kappa f^{\prime}\left(  R\right)  }\left[
V(\mathcal{P})+2g^{ij}\left(  \sqrt{g}f^{\prime}\left(  R\right)  \right)
_{\mid ij}\right]  =-\frac{\Lambda_{c}}{\kappa}\int_{\Sigma}d^{3}x\sqrt{g}.
\end{equation}
The term%
\begin{equation}
\frac{1}{\kappa}\int_{\Sigma}d^{3}x\frac{1}{f^{\prime}\left(  R\right)
}g^{ij}\left(  \sqrt{g}f^{\prime}\left(  R\right)  \right)  _{\mid ij}%
\end{equation}
appears to be a three-divergence and therefore will not contribute to the
computation. The remaining equation simplifies into%
\[
\int_{\Sigma}d^{3}x\left\{  \left[  \left(  2\kappa\right)  G_{ijkl}\pi
^{ij}\pi^{kl}{}-\frac{\sqrt{g}}{2\kappa}{}^{\left(  3\right)  }R\right]
+\left(  2\kappa\right)  \left[  G_{ijkl}\pi^{ij}\pi^{kl}+\frac{\pi^{2}}%
{4}\right]  \frac{2\left(  f^{\prime}\left(  R\right)  -1\right)  }{f^{\prime
}\left(  R\right)  }\right.
\]%
\begin{equation}
\left.  +\frac{V(\mathcal{P})}{2\kappa f^{\prime}\left(  R\right)  }\right\}
=-\frac{\Lambda_{c}}{\kappa}\int_{\Sigma}d^{3}x\sqrt{g}.\label{GWDW}%
\end{equation}
By a canonical procedure of quantization, we want to obtain the vacuum state
of a generic $f(R)$ theory. By repeating the same procedure for the
generalized WDW equation Eq.$\left(  \ref{GWDW}\right)  $, we obtain%
\[
\frac{1}{V}\frac{\left\langle \Psi\left\vert \int_{\Sigma}d^{3}x\left[
\hat{\Lambda}_{\Sigma}^{\left(  2\right)  }\right]  \right\vert \Psi
\right\rangle }{\left\langle \Psi|\Psi\right\rangle }+\frac{2\kappa}{V}%
\frac{2\left(  f^{\prime}\left(  R\right)  -1\right)  }{f^{\prime}\left(
R\right)  }\frac{\left\langle \Psi\left\vert \int_{\Sigma}d^{3}x\left[
G_{ijkl}\pi^{ij}\pi^{kl}+\pi^{2}/4\right]  \right\vert \Psi\right\rangle
}{\left\langle \Psi|\Psi\right\rangle }%
\]%
\begin{equation}
+\frac{1}{V}\frac{\left\langle \Psi\left\vert \int_{\Sigma}d^{3}%
xV(\mathcal{P})/\left(  2\kappa f^{\prime}\left(  R\right)  \right)
\right\vert \Psi\right\rangle }{\left\langle \Psi|\Psi\right\rangle }%
=-\frac{\Lambda_{c}}{\kappa}.\label{GWDW1}%
\end{equation}
From Eq.$\left(  \ref{GWDW1}\right)  $, we can define a \textquotedblleft%
\textit{modified}\textquotedblright\ $\hat{\Lambda}_{\Sigma}^{\left(
2\right)  }$ operator which includes $f^{\prime}\left(  R\right)  $. Thus, we
obtain%
\[
\frac{\left\langle \Psi\left\vert \int_{\Sigma}d^{3}x\left[  \hat{\Lambda
}_{\Sigma,f\left(  R\right)  }^{\left(  2\right)  }\right]  \right\vert
\Psi\right\rangle }{\left\langle \Psi|\Psi\right\rangle }+\frac{\kappa}%
{V}\frac{\left(  f^{\prime}\left(  R\right)  -1\right)  }{f^{\prime}\left(
R\right)  }\frac{\left\langle \Psi\left\vert \int_{\Sigma}d^{3}x\left[
\pi^{2}\right]  \right\vert \Psi\right\rangle }{\left\langle \Psi
|\Psi\right\rangle }%
\]%
\begin{equation}
+\frac{1}{V}\frac{\left\langle \Psi\left\vert \int_{\Sigma}d^{3}%
x\frac{V(\mathcal{P})}{2\kappa f^{\prime}\left(  R\right)  }\right\vert
\Psi\right\rangle }{\left\langle \Psi|\Psi\right\rangle }=-\frac{\Lambda_{c}%
}{\kappa},\label{GWDW2}%
\end{equation}
where%
\begin{equation}
\hat{\Lambda}_{\Sigma,f\left(  R\right)  }^{\left(  2\right)  }=\left(
2\kappa\right)  h\left(  R\right)  G_{ijkl}\pi^{ij}\pi^{kl}-\frac{\sqrt{g}%
}{2\kappa}\ ^{3}\!R^{lin},
\end{equation}
with%
\begin{equation}
h\left(  R\right)  =1+\frac{2\left[  f^{\prime}\left(  R\right)  -1\right]
}{f^{\prime}\left(  R\right)  }\label{h(R)}%
\end{equation}
and where $^{3}R^{lin}$ is the linearized scalar curvature whose expression is
shown in square brackets of Eq.$\left(  \ref{rexp}\right)  $. Note that when
$f\left(  R\right)  =R$, consistently it is $h\left(  R\right)  =1$. From
Eq.$\left(  \ref{GWDW2}\right)  $, we redefine $\Lambda_{c}$%
\begin{equation}
\Lambda_{c}^{\prime}=\Lambda_{c}+\frac{1}{2V}\frac{\left\langle \Psi\left\vert
\int_{\Sigma}d^{3}x\frac{V(\mathcal{P})}{f^{\prime}\left(  R\right)
}\right\vert \Psi\right\rangle }{\left\langle \Psi|\Psi\right\rangle }%
=\Lambda_{c}+\frac{1}{2V}\int_{\Sigma}d^{3}x\sqrt{g}\frac{Rf^{\prime}\left(
R\right)  -f\left(  R\right)  }{f^{\prime}\left(  R\right)  }%
,\label{NewLambda}%
\end{equation}
where we have explicitly used the definition of $V(\mathcal{P})$. In the same
spirit of the previous section, we restrict the analysis to the contribution
of physical degrees of freedom, namely TT tensors.\footnote{For a complete
derivation of the effective action for a $f\left(  R\right)  $ theory, see
Ref.\cite{CENOZ}.}\footnote{By a canonical decomposition of the gauge part
$\xi_{a}$ into a transverse part $\xi_{a}^{T}$ with $\nabla^{a}\xi_{a}^{T}=0$
and a longitudinal part $\xi_{a}^{\parallel}$ with $\xi_{a}^{\parallel}%
=\nabla_{a}\psi$, it is possible to show that most of the contribution comes
from the longitudinal part (scalar). Evidence against scalar perturbation
contribution in a Schwarzschild background has been discussed in
Ref.\cite{Remo1}.}. Thus Eq.$\left(  \ref{lambdamu0}\right)  $ becomes%
\begin{equation}
\frac{\Lambda_{0}^{\prime}\left(  \mu_{0},r\right)  }{8\pi G}=-\frac{m_{0}%
^{4}\left(  M\right)  }{128\pi^{2}}\left[  \ln\left(  \frac{m_{0}^{2}\left(
M\right)  }{4\mu_{0}^{2}}\right)  +\frac{1}{2}\right]  .\label{NewLambda'}%
\end{equation}
Now, we compute the maximum of $\Lambda_{0}^{\prime}$, by setting $x=m_{0}%
^{2}\left(  M\right)  /4\mu_{0}^{2}$. Thus $\Lambda_{0}^{\prime}$ becomes%
\begin{equation}
\Lambda_{0}^{\prime}\left(  \mu_{0},x\right)  =-\frac{G\mu_{0}^{4}}{\pi}%
x^{2}\left[  \ln\left(  x\right)  +\frac{1}{2}\right]  .
\end{equation}
As a function of $x$, $\Lambda_{0}\left(  \mu_{0},x\right)  $ vanishes for
$x=0$ and $x=\exp\left(  -\frac{1}{2}\right)  $ and when $x\in\left[
0,\exp\left(  -\frac{1}{2}\right)  \right]  $, $\Lambda_{0}^{\prime}\left(
\mu_{0},x\right)  \geq0$. It has a maximum for $\bar{x}=1/e$ equivalent to
$m_{0}^{2}\left(  M\right)  =4\mu_{0}^{2}/e$ and its value is%
\begin{equation}
\Lambda_{0}^{\prime}\left(  \mu_{0},\bar{x}\right)  =\frac{G\mu_{0}^{4}}{2\pi
e^{2}}%
\end{equation}
or%
\begin{equation}
\frac{1}{\sqrt{h\left(  R\right)  }}\left[  \Lambda_{0}\left(  \mu_{0},\bar
{x}\right)  +\frac{1}{2V}\int_{\Sigma}d^{3}x\sqrt{g}\frac{Rf^{\prime}\left(
R\right)  -f\left(  R\right)  }{f^{\prime}\left(  R\right)  }\right]
=\frac{G\mu_{0}^{4}}{2\pi e^{2}}.
\end{equation}
Isolating $\Lambda_{0}\left(  \mu_{0},\bar{x}\right)  $, we get%
\begin{equation}
\Lambda_{0}\left(  \mu_{0},\bar{x}\right)  =\sqrt{h\left(  R\right)  }%
\frac{G\mu_{0}^{4}}{2\pi e^{2}}-\frac{1}{2V}\int_{\Sigma}d^{3}x\sqrt{g}%
\frac{Rf^{\prime}\left(  R\right)  -f\left(  R\right)  }{f^{\prime}\left(
R\right)  }.
\end{equation}
Note that $\Lambda_{0}\left(  \mu_{0},\bar{x}\right)  $ can be set to zero
when%
\begin{equation}
\sqrt{h\left(  R\right)  }\frac{G\mu_{0}^{4}}{2\pi e^{2}}=\frac{1}{2V}%
\int_{\Sigma}d^{3}x\sqrt{g}\frac{Rf^{\prime}\left(  R\right)  -f\left(
R\right)  }{f^{\prime}\left(  R\right)  }.\label{lambda0_fin}%
\end{equation}
Let us see what happens when $f\left(  R\right)  =\exp\left(  -\alpha
R\right)  $. This choice is simply suggested by the regularity of the function
at every scale. In this case, Eq.$\left(  \ref{lambda0_fin}\right)  $ becomes%
\begin{equation}
\sqrt{\frac{3\alpha\exp\left(  -\alpha R\right)  +2}{\alpha\exp\left(  -\alpha
R\right)  }}\frac{G\mu_{0}^{4}}{\pi e^{2}}=\frac{1}{\alpha V}\int_{\Sigma
}d^{3}x\sqrt{g}\left(  1+\alpha R\right)  .
\end{equation}
For Schwarzschild, it is $R=0$, and by setting $\alpha=G$, we have the
relation%
\begin{equation}
\mu_{0}^{4}=\frac{\pi e^{2}}{G}\sqrt{\frac{1}{\left(  3G+2\right)  G}}.
\end{equation}
Like the case of $f\left(  R\right)  =R$, the purpose of this calculation is
related to Eq.$\left(  \ref{WDW2}\right)  $, which can be applied to different
backgrounds, i.e. de Sitter, Schwarzschild-de Sitter, etc. The result of this
process will be a spectrum built in terms of vacuum fluctuations. Once the
\textquotedblleft\textit{Ground State}\textquotedblright\ of this spectrum
will be identified, then there will be a chance to have an approach for
explaining the so-called \textit{Dark Energy}. Note that in this approach,
there is no evolution in time, so some pathologies of the type
\textquotedblleft\textit{Big Rip}\textquotedblright\ do not come into play at
this stage.


\section*{References}

\begin{thebibliography}{99}                                                                                               %


\bibitem {Lambda}For a pioneering review on this problem see S. Weinberg,
\textsl{Rev. Mod. Phys. }\textbf{61}, 1 (1989). For more recent and detailed
reviews see V. Sahni and A. Starobinsky, \textsl{Int. J. Mod. Phys.}
\textbf{D} \textbf{9}, 373 (2000), astro-ph/9904398; N. Straumann, \textit{The
history of the cosmological constant problem} gr-qc/0208027; T.Padmanabhan,
\textsl{Phys.Rept.} \textbf{380}, 235 (2003), hep-th/0212290.

\bibitem {De Witt}B. S. DeWitt, \textsl{Phys. Rev.} \textbf{160}, 1113 (1967).

\bibitem {Remo}R.Garattini, \textsl{J. Phys. A }\textbf{39}, 6393 (2006);
gr-qc/0510061. R. Garattini, \textsl{J.Phys.Conf.Ser}. \textbf{33}, 215
(2006); gr-qc/0510062.

\bibitem {Regge Wheeler}T. Regge and J. A. Wheeler, \textsl{Phys. Rev.
}\textbf{108}, 1063 (1957).

\bibitem {tHooft}G. 't Hooft, \textsl{Nucl. Phys.} \textbf{B} \textbf{256},
727 (1985).

\bibitem {RGeq}J.Perez-Mercader and S.D. Odintsov, \textsl{Int. J. Mod. Phys.}
\textbf{D} \textbf{1}, 401 (1992). I.O. Cherednikov, \textsl{Acta Physica
Slovaca}, \textbf{52}, (2002), 221. I.O. Cherednikov, \textsl{Acta Phys.
Polon.} \textbf{B} \textbf{35}, 1607 (2004). M. Bordag, U. Mohideen and V.M.
Mostepanenko, \textsl{Phys. Rep.} \textbf{353}, 1 (2001). Inclusion of
non-perturbative effects, namely beyond one-loop, in de Sitter Quantum Gravity
have been discussed in S. Falkenberg and S. D. Odintsov, \textsl{Int. J. Mod.
Phys. }\textbf{A} \textbf{13}, 607 (1998); hep-th 9612019.

\bibitem {CF}S. Capozziello and M. Francaviglia, \textit{Extended Theories of
Gravity and their Cosmological and Astrophysical Applications}, arXiv:0706.1146.

\bibitem {NO}S. Nojiri and S. D. Odintsov, \textsl{Int.J.Geom.Meth.Mod.Phys}.
\textbf{4}, 115 (2007); hep-th/0601213.

\bibitem {CG}S. Capozziello and R. Garattini, \textsl{Class.Quant.Grav.}
\textbf{24}, 1627 (2007); gr-qc/0702075.

\bibitem {Querella}L. Querella, \textit{Variational Principles and
Cosmological Models in Higher-Order Gravity} - Ph.D. Thesis. gr-qc/9902044.

\bibitem {CENOZ}G. Cognola, E. Elizalde, S. Nojiri, S. D. Odintsov and S.
Zerbini, \textsl{JCAP} \textbf{0502} 10 (2005); hep-th/0501096.

\bibitem {Remo1}R. Garattini, \textsl{TSPU Vestnik} \textbf{44} \textbf{N7},
72 (2004); gr-qc/0409016.

\bibitem {AbNoOd}M.C.B. Abdalla, S. Nojiri and S. D. Odintsov,
\textsl{Class.Quant.Grav}. \textbf{22, }L35 (2005); hep-th/0409177.
\end{thebibliography}
\end{document}